\documentclass[twocolumn,aps,pra]{revtex4-1}
\usepackage{amsmath}
\usepackage{graphicx}
\usepackage[latin1]{inputenc}
\usepackage{pstricks}
\usepackage{pst-node}
\usepackage{pst-coil}
\usepackage{pst-grad}
\usepackage{multido}
\usepackage{dcolumn}% Align table columns on decimal point
\usepackage{bm}% bold math

\begin{document}

\title{Microscopic dynamical Casimir effect}

\author{Reinaldo de Melo e Souza}
\affiliation{Instituto de F\'isica, Universidade Federal Fluminense, CP 24210-346, 24210-346 Niter\'oi, Rio de Janeiro, Brazil}
\author{Fran\c cois Impens}
\author{Paulo A. Maia Neto}
\affiliation{Instituto de F\'isica, Universidade Federal do Rio de Janeiro,  CP 68528, 21941-909 Rio de Janeiro,
Rio de Janeiro, Brazil}

\date{\today}

\begin{abstract}
We consider an atom in its ground state undergoing a non-relativistic oscillation in free space. The interaction with the electromagnetic quantum vacuum leads to two 
effects to leading order in perturbation theory. When the mechanical frequency is larger than the 
atomic transition frequency, the dominant effect is 
the motion-induced transition to an excited state with the emission of a photon carrying the excess energy.  
We compute 
the angular distribution of emitted photons and the excitation rate. 
On the other hand, when the mechanical frequency is smaller than  the transition frequency, the leading-order effect is the 
 parametric emission of photon pairs, which constitutes the microscopic counterpart of the dynamical Casimir effect. 
 We discuss the properties of the microscopic dynamical Casimir effect and build a connection with the photon production by an oscillating macroscopic 
 metallic mirror. 
 
\end{abstract}

%\pacs{34.35.+a,31.30.J-, 34.10.+x, 34.20.-b, 34.20.Gj, 31.15.am}

\maketitle
%
%
%%%%%%%%%%%%%%%%%%%%%%%%%%%%%%%%%%%%%%%%%%%%%%%%%%%%%%%%%%%%%%%%%%%%%%%%%%%%%%%%%%%%
%
%

\section{Introduction}

One of the cornerstones of classical electrodynamics is that accelerated point charges emit radiation. Within the realm of quantum mechanics, one might  expect
that a ground-state atom undergoing an accelerated motion could also produce radiation.  
Neutral macroscopic bodies in  non-uniform motion are predicted to emit photons out of the quantum vacuum state, an elusive effect that has so far defied experimental verification, and which is known as the dynamical Casimir effect (DCE) (for reviews see \cite{Dodonov2010,Dalvit2011}). 
Several geometries~\cite{Barton96B,Mundarain98,Kardar99,MaiaNeto05,Crocce05,Manjavacas2010} and
material models \cite{Barton95,Gutig98,Miri07}  have been analyzed. 
DCE is greatly enhanced by making use of a cavity 
resonance~\cite{Moore70,Castagnino84,Dodonov90,Jaekel92B,Lambrecht96,Dalvit98,Schaller02,Macri17}.
Analog models have been proposed \cite{Yablonovitch1989,Braggio2005,Johansson09,Lombardo16,Carusotto17} and realized experimentally \cite{Wilson2011}.
More recently, several applications in quantum information have been investigated \cite{sabin,sabin2,stassi,feli}.

Standard treatments of DCE are usually based on boundary conditions (or more generally scattering matrices) for 
 scalar  \cite{Fulling76,Ford82,Jaekel92A,Alves03,Rego13}  
or vector field operators \cite{MaiaNeto94,MaiaNeto96,Maghrebi2013} satisfying 
 the macroscopic Maxwell equations. 
In this paper, we investigate the  microscopic origin of DCE, by considering a ground-state atom undergoing a center-of-mass oscillation.
Instead of boundary conditions or scattering matrices, 
our approach builds on standard quantum optical Hamiltonian treatments for the coupling between  an atom 
in a highly excited external state of an atom trap and the 
  electromagnetic vacuum state. 
  In the same spirit of the Ewald-Oseen microscopic approach \cite{Fearn1996} to classical electrodynamics, 
  our main purpose is to gain insight into the physics of the DCE at the more fundamental atomic level and identify possible 
  universal features of this effect.

We consider a standard inertial frame in which the atom oscillates. However, our results can be reinterpreted in terms of  a co-moving frame.
The key point is that the vacuum state of a quantum field is in general not invariant under a transformation to a non-inertial frame.
For instance, in the Unruh effect, the vacuum is seen as a thermal field by an  observer with uniform proper acceleration \cite{Davis75,Unruh76}
(for a review see \cite{Crispino2008}).  

Moving atoms provide a particularly illuminating example in connection with the Unruh effect, since they behave as local probes of the quantum field. 
In the specific Unruh's scenario of uniform proper acceleration, 
the excitation of an internal state of a point-like detector~\cite{Unruh84,Ginzburg1987} or atom~\cite{Audretsch1994} coupled to a scalar field was analyzed in detail.
More recently, the interplay between entaglement and the Unruh effect for a two-atom system has been investigated \cite{Menezes2016}. 
Most  theoretical works address the specific case of a constant proper acceleration, in  which case no radiation is produced~\cite{Raine1991,Hu2004}. 

In contrast, a ground-state atom oscillating at a prescribed frequency does produce radiation, as discussed in this paper. 
By defining a mechanical frequency scale, we are able to develop a physical picture based on the principle of energy conservation and on the analogy with 
standard nonlinear optical effects. More general motions can be considered by generalizing our formalism to the case in which the motion contains 
different Fourier components. 
Considering an harmonic motion also allows us to build a direct comparison with the results for  oscillating 
planar plates \cite{MaiaNeto96} and
spheres \cite{MaiaNeto1993}, thus providing 
 insight into the microscopic origin of the DCE.

When the external oscillation frequency $\omega_{\rm cm}$ is larger than the atomic internal transition frequency, 
we show that the leading-order effect is the motion-induced transition to an excited internal state, 
with the emission of a single photon carrying the excess energy.  
The opposite limit, with $\omega_{\rm cm}$ much smaller than the transition frequencies, is more common for real atom traps. 
In this case, the external motion is quasi-static with respect to the internal degrees of freedom, but not with respect to the electromagnetic field modes with frequencies 
smaller than $\omega_{\rm cm}.$ 
Therefore, no real internal transition takes place, but the low-frequency field modes are parametrically excited by the DCE emission of photon pairs.

Here we calculate the DCE angular and frequency spectra for an atom in free space.
Atoms in motion  in the vicinity of a material surface gives rise to a variety of additional interesting effects, including 
 nonlocal~\cite{Impens2013, Impens2014}, non-additive~\cite{Impens2013B} and geometric \cite{Lombardo2017} phases, decoherence 
 of the internal degrees of freedom~\cite{Farias2016}, 
corrections to the Casimir-Polder interaction \cite{Hu2004,Scheel2009} and nonequilibrium forces \cite{Behunin2011}. 

This paper is organized as follows. The next section is dedicated to the motion-induced excitation effect involving a one-photon process. 
 In section III we  develop the theory of the microscopic DCE and consider in detail the corresponding two-photon emission process. 
Final remarks are presented in section IV.
  
\section{Motion-induced excitation \label{mie}}

Our system is composed by a ground-state atom undergoing a prescribed oscillation in an harmonic trap. We consider a regime where the external atomic motion can be treated classically. Our model approximates the case of semiclassical coherent wave-packets  in 
   magnetic or optical atom traps~\cite{FootNoteSectionIIclassical}. In order to simplify the notation, we consider in this section a two-level atomic model. 
   Details of the derivation are presented in Appendix B for the more general case of a 
 multi-level atom. 

To first order in perturbation theory, the atomic external motion might induce a transition to the excited state, accompanied by the emission of a single photon containing the excess energy, as illustrated by Fig.~1a. 
Such process is the analog of
 of the Stokes Raman effect with 
 the center-of-mass motion playing the role of the pump field. 
  Instead of  inelastic Raman scattering, 
 we have photon emission out of the vacuum field state.
We show that the corresponding angular distribution is in general anisotropic, and its shape is determined by the ratio between the mechanical frequency and the transition frequency.  Other analogies can be proposed. For instance, in the ionization process by monochromatic radiation, the ionized electron is the analog of the emitted photon, while the incident radiation plays the role of the center-of-mass motion.

We model our system by the Hamiltonian $H=H_A+H_F+H_{\rm int}$, where $H_A$ stands for the internal atomic degrees of freedom, $H_F$ for the free electromagnetic field 
 and  $H_{\rm int}$ describes the atom-field interaction.  In the dipole approximation, the interaction Hamiltonian assumes the following form
\begin{equation}
H_{\rm int}(t)=-\mathbf{d}\cdot\left[\mathbf{E}(\mathbf{r}(t))+\frac{\mathbf{v}(t)}{c}\times\mathbf{B}(\mathbf{r}(t))\right] , \label{hint}
\end{equation}
where $\mathbf{r}(t)$ is the atomic center of mass position  and $\mathbf{v}(t)=d\mathbf{r}(t)/dt$ the associated velocity. 
The operators in the interaction picture
$\mathbf{E}$ and $\mathbf{B}$ 
represent the electric and magnetic
field, respectively, whereas 
 $\mathbf{d}$ denotes the atomic electric dipole. 
We assume the motion to be non-relativistic so that $v(t) \ll c$ at all times. The second term in the right-hand-side of Eq.~(\ref{hint})
stands for the R\"ontgen contribution and is crucial to assure
 Lorentz covariance to first order in $v/c$~\cite{Wilkens1994,baxter}. 

Initially, the atom is in the ground state and the electromagnetic field is in the vacuum state. We investigate the population of one-photon states resulting from the atomic motion. 
We use the notation  $|s , 1_{\mathbf{k}\lambda}\rangle \equiv |s\rangle\otimes|1_{\mathbf{k}\lambda}\rangle,$ 
where $s$ represents the internal ground ($g$) or excited ($e$) state, while $|1_{\mathbf{k}\lambda}\rangle$ is the one-photon field state with wave-vector
 $\mathbf{k}$ and  polarization $\lambda.$ 
 The  probability for photon emission 
 after a duration $T$ is given by standard first-order time-dependent perturbation theory:
\begin{equation} 
p_{\mathbf{k}\lambda} = \frac{1}{\hbar^2}\Bigg|\int\limits_{0}^{T}\langle \,1_{\mathbf{k}\lambda}, \, e  | H_{\rm int}(t)|g, \, 0\rangle dt\Bigg|^2 \, . \label{psk}
\end{equation}
Note that the absence of permanent  dipole moment for the ground state  implies that the one-photon emission process only occurs by concomitantly exciting the atom. 

\begin{figure}
\includegraphics[scale=0.24]{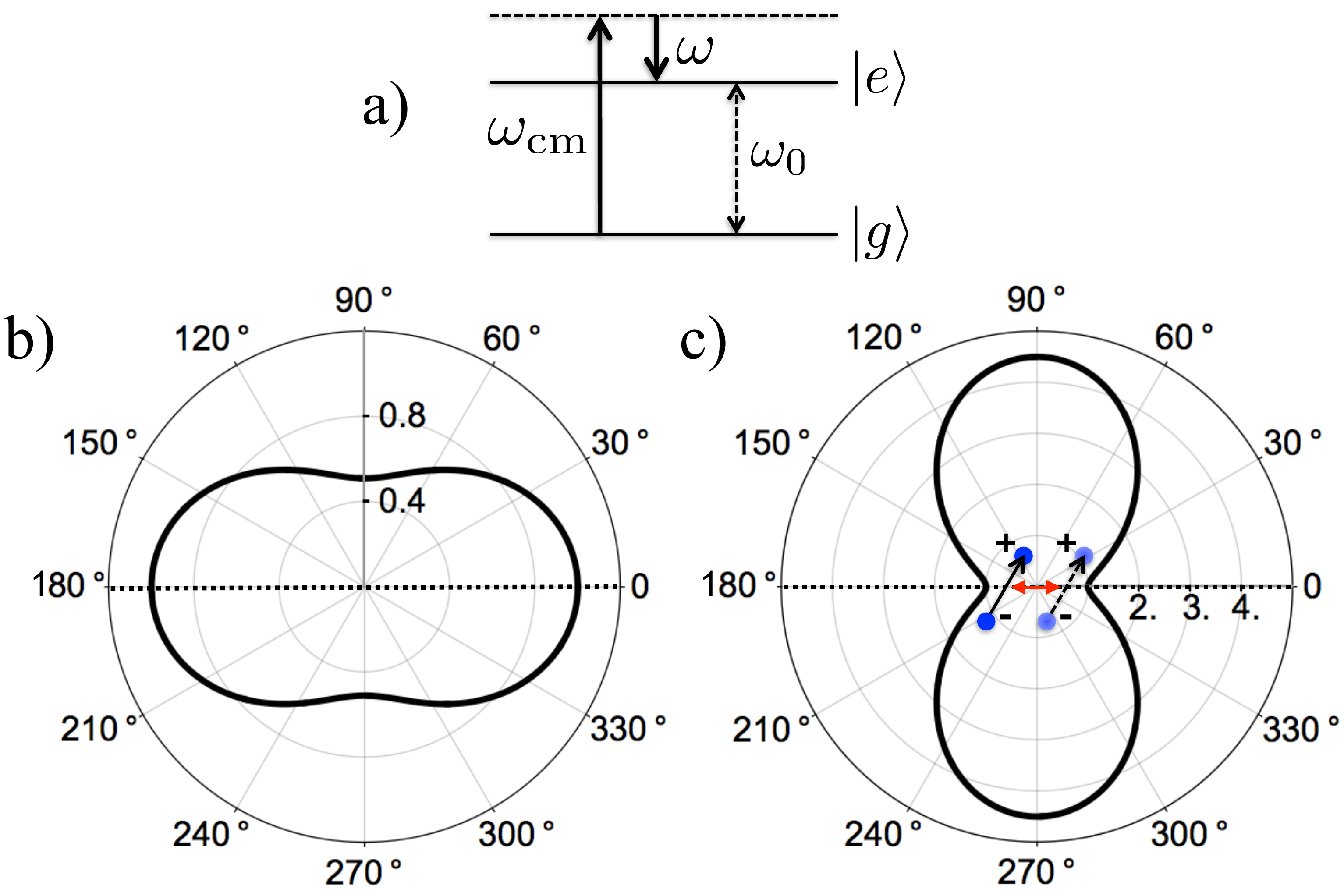}
\caption{ (a) Energy level diagram for the motion-induced excitation showing the internal ($\omega_0$), external ($\omega_{\rm cm}$) and photon  ($\omega$) frequencies. 
(b,c) Angular distribution of the light emitted through motion-induced excitation for (b) $\omega_{\rm cm}=1.01\,\omega_{0}$ and 
(c) $\omega_{\rm cm}=3\,\omega_{0}$. The distributions are normalized by the value of the emission rate per unit solid angle along the direction of the external motion, which is indicated by a horizontal dashed line. 
}
\label{excitation}
\end{figure}

When computing (\ref{psk}), we take 
\begin{equation}\label{cm}
\mathbf{r}(t)=\boldsymbol{a}\cos(\omega_{\rm cm} t)
\end{equation}
in Eq.~(\ref{hint}).   We also define the maximum external velocity $v_m= \omega_{\rm cm} a.$
We consider long interaction times $ T\gg 1/\omega_0,$ where $\omega_0$ is the atomic transition frequency.  This corresponds to a stationary regime 
in which only resonant processes contribute. Although we describe the external motion as a classical prescribed trajectory,  Eq.~(\ref{cm}) defines 
the energy quantum  $\hbar\omega_{\rm cm}$ as usual in time-dependent  perturbation theory with a sinusoidal perturbation Hamiltonian \cite{Landau}. 
Thus,  the emitted spectrum only contains a single frequency $\omega_{\mathbf{k}}=c\,|\mathbf{k}|=\omega_{\rm cm}-\omega_{0}$ as illustrated in Figure~\ref{excitation}a, provided that
 $\omega_{\rm cm}>\omega_{0}$. Since $\omega_{\mathbf{k}}<\omega_{\rm cm},$ the external amplitude $a$ is much smaller than the relevant field wavelengths: $\omega_{\mathbf{k}} a/c < v_m/c\ll 1,$ allowing us to expand 
 the electromagnetic fields in Eq.~(\ref{hint}) to linear order in $a.$ 
 
 The number of emitted photons is proportional to the duration $T$ in the stationary regime, enabling the definition of a photon emission rate $\Gamma_{\rm MIE}$ describing  the motion-induced excitation (MIE).
We 
express our results in 
 terms of the spontaneous emission rate $\Gamma_0=  |\langle e | \mathbf{d}(0) | g \rangle|^2 \omega_0^3 / (6 \pi \hbar c^3)$ of the two-level atom at rest.
  We first compute the  angular distribution of photons, namely the number of photons emitted 
  per unit of solid angle and per unit time (see Appendix A for details): 
\begin{eqnarray}
\frac{d \Gamma_{\rm MIE}}{d\Omega_{\mathbf{\hat{k}}}}&=&\frac{\Gamma_0 v_m^2}{4 c^2 }  \Theta(\omega_{\rm cm}-\omega_{0}) \left(\frac {\omega_{\rm cm}} {\omega_{0}}-1 \right)^3 \cr\cr
&& \times \left[ 2 \left( \frac {\omega_{0}} {\omega_{\rm cm}} \right)^2 (\mathbf{\hat{k}}\cdot \hat{\boldsymbol{a}})^2 + (\mathbf{\hat{k}}\times \hat{\boldsymbol{a}})^2 \right] 	\, .  \nonumber\\ 
\label{dndomega}
\end{eqnarray}
We have introduced the Heaviside function defined as $\Theta(x)=1$ if $x \geq 0$ and $ \Theta(x)=0$ if $x<0,$ as well as the unit vectors $\hat{\mathbf{k}}$ and $\hat{\boldsymbol{a}}$ along the directions of the photon emission and of the atomic motion respectively. 

 Eq.~\eqref{dndomega} shows that the angular distribution receives two separate contributions associated to the projections of the wave-vector parallel or perpendicular to  the direction of motion. 
 For $\omega_{\rm cm}=\sqrt{2}\omega_0,$ these two projections contribute with  equal weights, yielding  an isotropic radiation in this case.
When  $\omega_{\rm cm}$ is smaller (larger) than $\sqrt{2}\omega_0,$ the angular distribution is maximum (minimum) along the direction of motion. 
The radiation emitted by
 motion-induced excitation can be  highly anisotropic, as illustrated 
 in Figs.~\ref{excitation}b and \ref{excitation}c.

We take a mechanical frequency  barely higher
 than the atomic transition frequency  in Fig.~\ref{excitation}b.  In this case,  the moving atom radiates nearly twice along the direction of motion as compared to the orthogonal direction. 
 In classical electrodynamics, emission by an accelerated point-like electric dipole along the direction of motion is also possible~\cite{heras}. However, no classical analogy is available when the frequency scales for the 
 atomic dipole fluctuations ($\omega_0$) and for the external motion ($\omega_{\rm cm}$) are comparable. 
 On the other hand, when $\omega_{\rm cm}\gg\omega_0,$ the distribution illustrated by Figure~\ref{excitation}c approaches a classical antenna-like angular distribution. In this limit, the slow dipole fluctuations may be neglected during the fast center of mass oscillation.
 The resulting radiation pattern may then be constructed by averaging the classical distribution over all possible 
atomic dipole orientations. This is illustrated by the inset of  Fig.~\ref{excitation}c, which suggests that the radiation field can be obtained by the superposition of the fields produced by the oscillating point charges with opposite signs.

For a multi-level atom, Eq.~(\ref{dndomega}) gives the contribution of each possible transition to the total angular distribution. Since the atomic excitation is
 accompanied by the emission of a single photon, we can obtain the excitation rate $\Gamma_{\rm MIE}$ for the transition of frequency $\omega_0$ by integrating 
 the r.h.s. of (\ref{dndomega}) over the solid angle:
\begin{eqnarray}
\frac{\Gamma_{\rm MIE}}{\Gamma_0} \! \! = \! \! \frac {2 v_m^2} {3 c^2} \Theta(\omega_{\rm cm}-\omega_{0}) \left(1+\frac  {\omega_{0}}{\omega_{\rm cm}}\right)^2 \! \!  \left( \frac {\omega_{\rm cm}} {\omega_{0}}-1 \right)^3  \label{onephotonemissionrate}
\end{eqnarray}
The excitation rate scales as $(v _m/c)^2$~\cite{FootnoteSecIIb} and is an increasing function of 
the center-of-mass  frequency $\omega_{\rm cm}.$ The frequency dependence results in part from 
 the density of field modes at frequency 
 $\omega=\omega_{\rm cm}-\omega_{0}$
  to which the ground-state atom is resonantly coupled through the atomic motion, given that the atom ends up in an excited state (see Fig.~1a). 
  Indeed, the larger the difference between $\omega_{\rm cm}$  and the $\omega_{0}$, the larger the density of vacuum modes accessible for the coupling through motion-induced excitation. Should the mechanical frequency be smaller than the transition frequency, no resonant process can take place to first order in the interaction. The corresponding contribution then vanishes, as indicated by the presence of the Heaviside function in
 (\ref{dndomega}) and   (\ref{onephotonemissionrate}). 
 However, 
 photon emission through  higher-order resonant processes may still occur in this case, as discussed in the next section.

\section{Microscopic dynamical Casimir effect}

In this section, 
we consider the microscopic DCE arising from a ground-state atom undergoing a mechanical oscillation. 
We assume that the external frequency is smaller than the smallest atomic transition frequency.
In this case, and
differently from the previous section, only virtual atomic excitations may occur up to second-order in the interaction~\cite{secondorder},
and the atom stays in the ground state at all times, as illustrated by Fig.~2a.
The atom-field interaction may then be described by an effective Hamiltonian 
obtained from the standard dipolar Hamiltonian through a unitary transformation~\cite{passante}:
\begin{equation}
\label{HamiltonianPassanteAtomicFrame}
H_{\rm eff}^{\rm rest}(\mathbf{r})=-\frac{1}{2}\sum_{\mathbf{k}\lambda}\alpha(\omega_{\mathbf{k}})\mathbf{E}_{\mathbf{k}\lambda}(\mathbf{r})\cdot\mathbf{E}(\mathbf{r}) \, ,
\end{equation}
written here for the instantaneous rest frame of the atom. In (\ref{HamiltonianPassanteAtomicFrame}), 
 $\alpha(\omega)$ stands for the atomic polarizability, given for freely rotating atoms by a sum over all possible excited states~\cite{thiru} 
\begin{equation}
\alpha(\omega)=(2 / 3 \hbar) \sum_{e} \omega_{eg}  |\langle e | \mathbf{d}(0) | g \rangle|^2 / ( 
\omega^2_{eg}-\omega^2) \, . \label{alpha}
\end{equation}
 $H_{\rm eff}^{\rm rest}$ is quadratic in the electric field and thus leads to the generation of photon pairs out of the vacuum state, as despicted in Fig.~2a and discussed in detail below. 

There are two main advantages in using Eq.~(\ref{HamiltonianPassanteAtomicFrame}) instead of the more standard dipolar Hamiltonian (\ref{hint}). First, virtual transitions are accounted for through the atomic polarizability so that $H_{\rm eff}^{\rm rest}$ does not operate on the internal atomic degrees of freedom -- it simply acts on the Hilbert space associated to the electromagnetic field. Second, the microscopic DCE is obtained already to first order of  perturbation theory, whereas 
a more involved second-order derivation would be required when using (\ref{hint}).

In order to obtain a description of the DCE in the laboratory frame, one must  Lorentz transform the electric field in Eq.~(\ref{HamiltonianPassanteAtomicFrame}).
We assume the external motion to be nonrelativistic and expand to first order in $v/c,$ leading to an 
 effective Hamiltonian in the laboratory frame  containing a R\"ontgen interaction term:
\begin{eqnarray}
H_{\rm eff}=&
H_{\rm eff}^{\rm rest}(\mathbf{r}(t))+\frac{\mathbf{v}(t)}{2c}\cdot 
\sum_{\mathbf{k}\lambda}&\alpha(\omega_{\mathbf{k}})\Bigg[\mathbf{E}_{\mathbf{k}\lambda}(\mathbf{r}(t))\times\mathbf{B}(\mathbf{r}(t)) \nonumber \\
&&-\mathbf{B}_{\mathbf{k}\lambda}(\mathbf{r}(t))\times \mathbf{E}(\mathbf{r}(t))
\Bigg], \label{hpassa}
\end{eqnarray}
with the field operators taken in the interaction picture.

We consider the sinusoidal motion (\ref{cm}) 
and the velocity is $\mathbf{v}(t)=-v_m\,\sin(\omega_{\rm cm}t)\, \boldsymbol{\hat{a}}.$ 
The field frequencies are bounded by $\omega_{\rm cm},$ allowing us to expand the rhs of (\ref{hpassa}) to first order in 
$\omega_{\mathbf{k}} a/c \ll 1$ as in   Sec.~II.
However, in contrast with the analysis of Sec.~II,  we assume that $\omega_{\mathbf{k}}<\omega_{\rm cm}\ll \omega_{eg},$ for all atomic internal transitions so that the relevant field modes are very slow in comparison with the atomic internal dynamics. 
As a consequence, we see from Eq.(\ref{alpha}) that we can approximate the dynamical polarizability by the static one, $\alpha(\omega_{\mathbf{k}})\approx \alpha(0)\equiv \alpha_0,$ leading to further simplification of (\ref{hpassa}).
We then find
\begin{eqnarray}
H_{\rm eff} (t)&\approx&  -\frac12 \alpha_0   \mathbf{E}^2+  V_{\alpha}\cos(\omega_{\rm cm} t)  + V_{\beta}\sin(\omega_{\rm cm} t) \label{Veff}  \\
V_{\alpha} &=&  -\frac12 \alpha_0\,  \boldsymbol{a}\cdot{\mathbf{\nabla}}\, \mathbf{E}^2 \\
V_{\beta}  &=&  - \frac12 \,\alpha_0  \, \frac{v_m}{c}\boldsymbol{\hat{a}} \cdot \left[ \mathbf{E}\times  \mathbf{B}- \mathbf{B} \times \mathbf{E} \right]
\end{eqnarray}
with all field operators taken at $\mathbf{r}=\mathbf{0}.$ The production of photon pairs results from the terms which depend explicitly on time, namely the ones proportional to $V_{\alpha}$ and $V_{\beta}$ in Eq.~(\ref{Veff}), with the latter accounting for the R\"ontgen contribution. 

It is insightful to build an analogy between $H_{\rm eff}$   and the Hamiltonian describing the emission of photon pairs by  spontaneous parametric down-conversion  in nonlinear crystals~\cite{Rubin94}.
 Both Hamiltonians are quadratic in the electric field. Here, the external oscillation plays the role devoted to the laser pump, and the atomic polarizability is the analog of the non-linear susceptibility. The quantum state of the light field resulting from the microscopic DCE 
  can be obtained using first-order time-dependent perturbation theory. Its generic decomposition 
in terms of two-photon states is given by
\begin{equation}
| \psi (t) \rangle = | 0 \rangle + \sum_{\mathbf{k}_1 \lambda_1 \mathbf{k}_2 \lambda_2} c_{\mathbf{k}_1 \lambda_1 \mathbf{k}_2  \lambda_2} (t)  | 1_{\mathbf{k}_1  \lambda_1} 1_{\mathbf{k}_2 \lambda_2} \rangle
\label{psit}
\end{equation}
We compute the two-photon amplitudes $c_{\mathbf{k}_1 \lambda_1 \mathbf{k}_2  \lambda_2} (t) $ to first order in the perturbation $ H_{\rm eff}$ and 
   take the rotating-wave approximation:
   \begin{eqnarray}
\label{eq:coeffsck1k2}
&&c_{\mathbf{k}_1 \lambda_1 \mathbf{k}_2  \lambda_2} (t) = -\frac {2 \pi  \alpha_0 v_m} {L^3 c} (\omega_1\omega_2)^{1/2}  e^{i\Delta\omega t/2} \,\frac{\sin(\Delta\omega\, t/2)}{\Delta\omega}
\nonumber \\
&\times &   \hat{\boldsymbol{a}} \cdot
\! \left[  \!  \frac {c} {\omega_{\rm cm}}\, \boldsymbol{\varepsilon}_{\mathbf{k}_1 \lambda_1} \! \cdot \! \boldsymbol{\varepsilon}_{\mathbf{k}_2 \lambda_2}\,\left(   {\mathbf{k}}_{1} + {\mathbf{k}}_{2}  \right)\right. \nonumber \\
 &+ & \left. ( \hat{\mathbf{k}}_{1}  \times  \boldsymbol{\varepsilon}_{\mathbf{k}_1 \lambda_1} )\times\boldsymbol{\varepsilon}_{\mathbf{k}_2 \lambda_2}  +( \hat{\mathbf{k}}_{2}  \times  \boldsymbol{\varepsilon}_{\mathbf{k}_2 \lambda_2}   )\times\boldsymbol{\varepsilon}_{\mathbf{k}_1 \lambda_1}\right]\nonumber 
\\
 \label{eq:coeffquantum}
\end{eqnarray}
where $\Delta\omega = \omega_1+\omega_2-\omega_{\rm cm}.$
We have introduced the shorthand notation 
$\omega_j\equiv \omega_{{\mathbf{k}}_j}$ and
the volume $L^3$ associated to the quantization of the electromagnetic field.

We now investigate the photon emission spectrum in the stationary regime. By taking the long time limit and squaring the coefficients given by 
 Eq.~(\ref{eq:coeffsck1k2}), one retrieves the Fermi golden rule for the probability of two-photon emission.
In this limit, energy conservation is enforced and the photon frequencies of the emitted pair satisfy (see Fig.~2a)
\begin{equation}
\omega_1+\omega_2=\omega_{\rm cm}.\label{sumfreqs}
\end{equation}
As a consequence, the frequencies are distributed in the range $0\le \omega \le \omega_{\rm cm}.$
The emission probability increases linearly with the interaction time $t,$  enabling the definition of a stationary radiation emission rate.
We first compute the angular spectrum representing the number of photons with polarization $\lambda$ emitted per unit of time, solid angle and frequency interval. For that purpose, we  sum over all possible wavevectors and polarizations for the accompanying photon, as detailed in Appendix B:
\begin{eqnarray}
\frac{d \Gamma^{(\lambda)}_{\rm \scriptscriptstyle DCE}}{d\omega d\Omega_{\mathbf{\hat{k}}}}(\omega,\theta)\!  &=&  \! \frac {(\alpha_0 v_m)^2 } {60\pi^2c^8}\omega^3(\omega_{\rm cm}-\omega)^3  f^{(\lambda)} \left( \frac {\omega} {\omega_{\rm cm}}, \theta \right)  \nonumber \\
\label{eq:angularspectrum}\\
 f^{({\rm TE})}(x,\theta)\! &=&\!  \left(1- x \right)^2 (5 \cos^2\theta+ 2) +5 x\label{TE}\\
 f^{({\rm TM})}(x,\theta)\!&=&\! (1-x)(1-6x) \cos^2\theta+(1-x)^2+5\nonumber,\\
 \label{TM}
\end{eqnarray}
where $\theta$ is the angle between the direction of photon emission $ \hat{\mathbf{k}}$ and the direction of motion $\hat{\boldsymbol{a}}.$
We have denoted
the polarization  with the electric field perpendicular to the plane defined by 
the unit vectors $ \hat{\mathbf{k}}$ and $\hat{\boldsymbol{a}}$
 as transverse electric (TE), and likewise for the 
 transverse magnetic (TM) one. 
 When $\theta=0, \pi,$  the angular spectra must be independent of polarization by symmetry, as can be verified from Eqs.~(\ref{eq:angularspectrum})-(\ref{TM}). 
 Our results also check a second symmetry property:  the angular spectra must be invariant when replacing $\theta\rightarrow \pi - \theta,$ 
since the two opposite directions $\hat{\boldsymbol{a}}$ and  $-\hat{\boldsymbol{a}}$ are equivalent for our harmonic motion when considering long interaction times.

\begin{figure}
\includegraphics[scale=0.25]{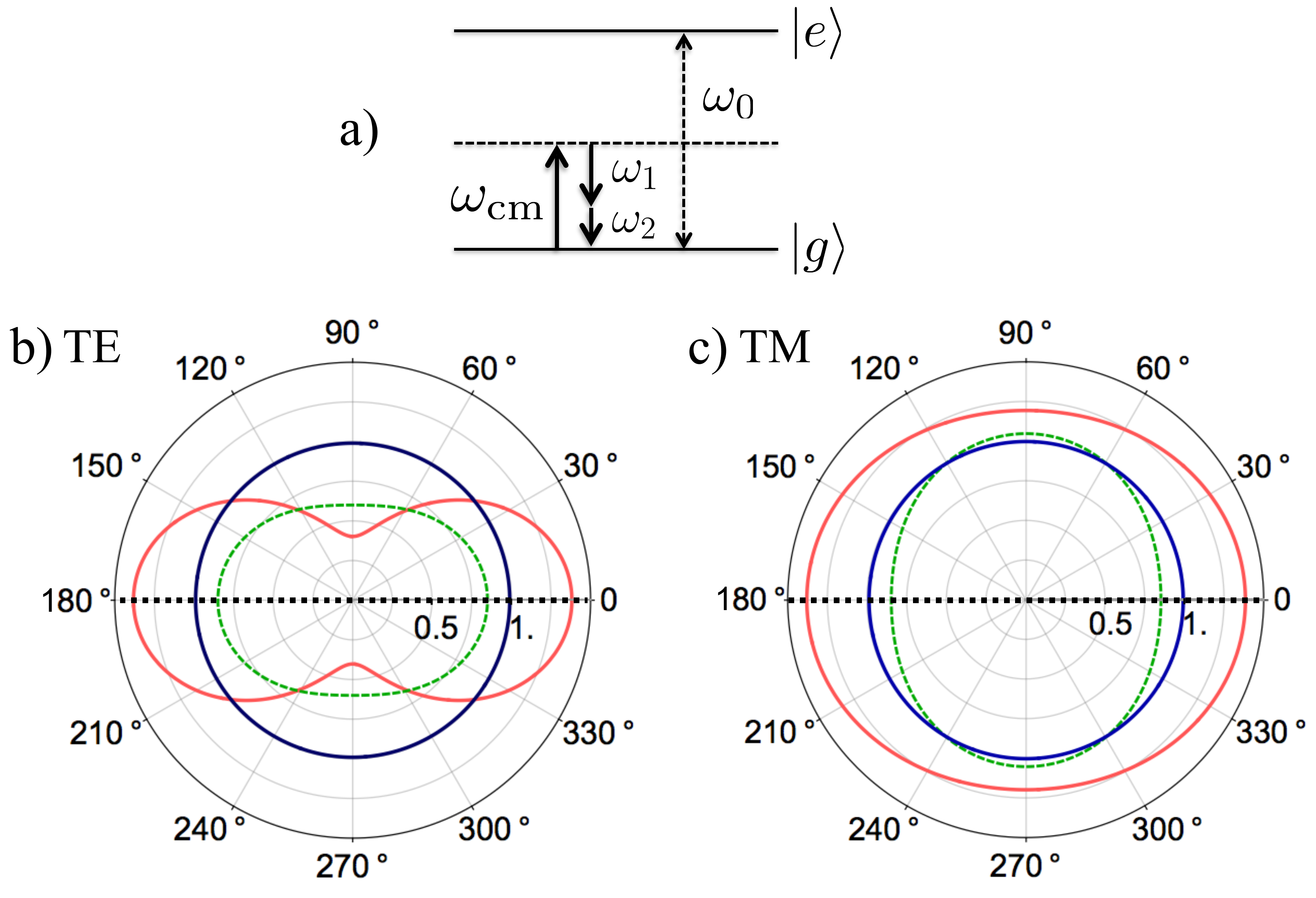}
\caption{
(a) Energy level diagram for the microscopic dynamical Casimir effect showing the internal ($\omega_0$), external ($\omega_{\rm cm}$) and photon frequencies ($\omega_1$ and $\omega_2$).
(b,c) DCE angular distributions for  (b) TE and (c) TM polarizations. 
The red (light gray), green (dashed line) and blue (dark gray) correspond to photon frequencies 
$\omega=0.01\, \omega_{\rm cm},$ $0.5\,\omega_{\rm cm}$ and $0.99\, \omega_{\rm cm},$ respectively.
The distributions are normalized by the value at $\omega=\omega_{\rm cm}$  along the direction of the external motion (horizontal dotted line).\label{distrangl}}
\end{figure}

The sign of the coefficient multiplying $\cos^2\theta$ in 
Eqs.~(\ref{TE}) and (\ref{TM}) determines the shape of the angular spectrum. For  TE polarization, the coefficient $5\left(1- \omega/\omega_{\rm cm} \right)^2$ is non-negative and hence the distribution is  elongated along the direction of motion, except at the upper frequency limit $\omega\rightarrow\omega_{\rm cm}.$
At this limit, 
 both TE and TM distributions become isotropic, but the  intensity  vanishes  as the local density of states becomes arbitrarily small.
The TM angular distribution   also favors emission close to the direction of motion for frequencies $\omega<\omega_{\rm cm}/6,$ but then becomes more elongated perpendicular to this direction for frequencies above $\omega_{\rm cm}/6,$
since the coefficient $(1-\omega/\omega_{\rm cm})(1-6\omega/\omega_{\rm cm})$ in (\ref{TM}) becomes negative in this case. 
Such properties are illustrated by 
figures~\ref{distrangl}b and \ref{distrangl}c for TE and TM polarizations, respectively. 

In short, the direction of motion is always a maximum of the TE distribution and a minimum of the TM one for most of the frequency range. 
 Since the two distributions coincide along this direction, this  observation suggests that there are more TM than TE emitted photons.
To be more quantitative, we first compute the 
TE and TM frequency spectra by a solid angle integration of Eq.~(\ref{eq:angularspectrum}):

\begin{eqnarray}
\frac{d \Gamma^{(\lambda)}_{\rm \scriptscriptstyle DCE}}{d\omega}(\omega)\!  &=&  \! \frac {(\alpha_0 v_m)^2 } {45 \pi c^8}
\omega^3(\omega_{\rm cm}-\omega)^3   F^{(\lambda)} \left(  \omega / \omega_{\rm cm} \right) \nonumber \\
\label{eq:spectrum}\\
 F^{({\rm TE})}(x) &=& 11x^2-7x+11\label{TEf}\\
 F^{({\rm TM})}(x)&=&   9x^2-13x+19
 \label{TMf}
\end{eqnarray}
in the range $0\le \omega\le \omega_{\rm cm}.$ The total frequency spectrum 
\[
\frac{d \Gamma_{\rm \scriptscriptstyle DCE}}{d\omega}(\omega) = \frac {2(\alpha_0 a)^2 } {3 \pi c^8} 
\omega^3(\omega_{\rm cm}-\omega)^3 \left[\omega_{\rm cm}^2-\frac23 \omega(\omega_{\rm cm}-\omega)\right]
\]
is invariant under the transformation $\omega  \rightarrow \omega_{\rm cm}-\omega.$ 
This is a direct consequence of the 
energy conservation condition for the emitted photon pair  given by 
Eq.~(\ref{sumfreqs}). Indeed, each photon emitted at frequency $\omega$  is accompanied by a twin emitted at frequency $\omega_{\rm cm}-\omega.$
The same property holds for the DCE with a macroscopic planar surface \cite{MaiaNeto96}. However, whereas 
for the latter the TE and TM spectra are also separately symmetric with respect to  $\omega=\omega_{\rm cm}/2,$ here 
the TE (TM) spectrum is slightly shifted towards frequencies larger (smaller) than $\omega_{\rm cm}/2.$ Such asymmetry arises from the emission of 
mixed TE-TM pairs, preferably with the TE twin emitted at the upper half of the frequency interval.

We obtain the total emission rates for each polarization  by
integrating (\ref{eq:spectrum}) over the frequency interval $[0,\omega_{\rm cm}].$ The resulting TM rate is larger than 
 the TE one by approximately $42\%.$ 
 The total rate is given by 
 $\Gamma_{\rm atom}=(23/5670\pi)  (\alpha_0 a)^2  \omega_{\rm cm}^9/c^8.$
The same frequency dependence can be found in a  different context, involving a macroscopic metallic sphere treated in terms of boundary conditions. 
 In fact, we can use the principle of energy conservation in order to derive  the total photon emission rate from the result for the vacuum dissipative force on an oscillating perfectly-refleting sphere obtained in Ref.~\cite{MaiaNeto1993}.
 When the sphere radius $R$ is much smaller than the typical field wavelength $\lambda\sim 2\pi c/\omega_{\rm cm},$
 we find $\Gamma_{\rm sphere}=(1/10368\pi^3)  (\alpha_{\rm sph} a)^2  \omega_{\rm cm}^9/c^8,$ where $\alpha_{\rm sph}=4\pi R^3$ is the electric polarizability of the metallic sphere \cite{comment_comparison}.
 
Such comparison between the total emission rates for an atom and a metallic sphere suggests that our microscopic approach is capable of explaining several
  features of the DCE for macroscopic bodies.  
In classical electrodynamics, physical insight is obtained by treating material media as a collection of dipoles, instead of employing the more standard
 macroscopic Maxwell equations and the corresponding boundary conditions, as often discussed in the context of the 
 Ewald-Oseen extinction theorem~\cite{Fearn1996}.

Here we propose to build the first steps of a similar construction concerning the DCE. In classical electrodynamics, 
the case of a material medium with a planar interface 
provides the most illustrative example for the comparison with the microscopic approach.
 For the DCE, Ref.~\cite{MaiaNeto96} presents a detailed macroscopic theory of the radiation emitted by an oscillating perfectly-reflecting planar interface. 
 Our results for a single atom already share some common features with the DCE by a planar interface: there are more TM than TE photons, and TE photons are preferably emited close to the direction of motion.
 In order 
 to bridge  the gap 
 between 
 \cite{MaiaNeto96} and
  our microscopic results, we consider that the material half-space, limited by a planar interface, is constituted of ground-state atoms oscillating in phase along the
   direction $\hat{\boldsymbol{a}} $ perpendicular to the interface.
Symmetry of  translation parallel to the interface implies that the
two photons of a given pair have the same polarization and 
 satisfy the condition 
 \begin{equation}\label{condition_planar_symmetry}
 \hat{\boldsymbol{a}} \times ({\bf k}_1+{\bf k}_2)=\mathbf{0},
 \end{equation}
in  addition to  energy conservation (\ref{sumfreqs}).
Accordingly, we assume that the emission amplitudes associated to different atoms interfere destructively 
 except for the propagation directions satisfying (\ref{condition_planar_symmetry}), and for all directions when considering mixed TE-TM pairs. 

\begin{figure}
\includegraphics[scale=0.43]{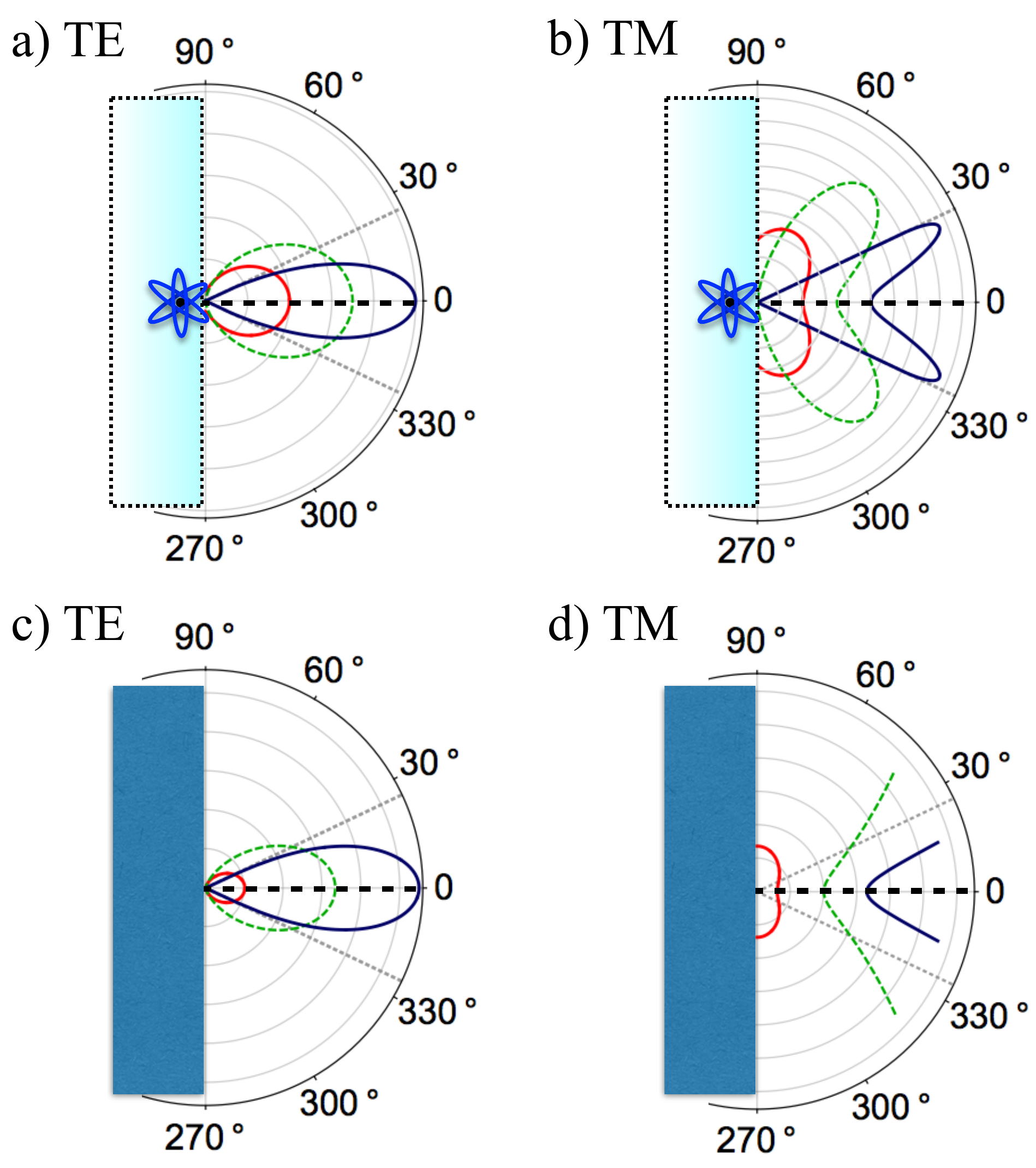}
\caption{ Comparison between the angular spectra arising from  the oscillation of a single atom
for (a) TE and (b) TM polarizations, with the spectra for an oscillating perfectly-reflecting mirror, also show for  (c) TE and  (d) TM polarizations. 
For the atomic case, we only consider photon pairs satisfying the constraint
(\ref{condition_planar_symmetry}) associated to the planar symmetry. 
The red (light gray), green (dashed gray) and blue (dark gray) correspond to photon frequencies 
$\omega=0.3\, \omega_{\rm cm},$ $0.5\, \omega_{\rm cm}$ and $0.7\, \omega_{\rm cm},$ respectively. In the last case, emission is restricted to the angular sector $\theta\le \arcsin(\omega_{\rm cm}/\omega-1)\approx 25^{\rm o}$
bounded by the dotted thin lines. Both atom (a,b) and mirror (c,d) oscillate along the direction indicated by the horizontal dashed black line. The angular distributions associated to different frequencies have been plotted using different (arbitrary) scales in (a,b).} \label{fig:connectionplane}
\end{figure}

We now compute the angular spectra from Eq.~(\ref{eq:coeffquantum}) by enforcing such symmetry conditions.  Then, a given 
${\bf k}_1$ and $\lambda_1$ determines a single possibility for the accompanying photon wave-vector ${\bf k}_2$ and polarization $\lambda_2=\lambda_1.$  The resulting angular distributions for TE and TM polarizations are sketched in panels (a) and (b) of Fig.~\ref{fig:connectionplane}, respectively.
 We also show the angular spectra for a perfectly-reflecting plane surface in panels (c) (TE) and (d) (TM) calculated in Ref.~\cite{MaiaNeto96}.  For frequencies in the upper half-interval $\omega_{\rm cm}/2 < \omega \le\omega_{\rm cm},$ Eqs.~
 (\ref{sumfreqs}) and 
 (\ref{condition_planar_symmetry}) jointly imply that emission is restricted to the angular sector around the  direction of motion given by 
 $\theta\le \theta_0=\arcsin(\omega_{\rm cm}/\omega-1).$ 
For the atomic case shown  in panel (b), the TM distribution develops a sharp peak near the boundary $\theta_0,$
 whereas 
 for the macroscopic perfect reflector shown in panel (d), the TM distribution diverges as $\theta\to\theta_0.$
The comparison between the  atomic and the  perfect reflector distributions indicates that the highly singular behavior of the latter at $\theta=\theta_0$ results from the 
unphysical assumption of perfect reflectivity.
 For all frequencies, the direction of motion  is a local minimum 
 for TM photons,  and a maximum for TE polarization, in both atomic and macroscopic cases. 
 
 Overall, Fig.~\ref{fig:connectionplane} shows that the
 main properties of the spectra for a plane perfectly-reflecting surface are already present at the atomic level 
when considering only photon pairs that do not violate the planar symmetry: TE photons are mostly emitted near the direction of motion, whereas TM photons are preferably emitted as far from this direction as allowed by conditions 
 (\ref{sumfreqs}) and  (\ref{condition_planar_symmetry}).

\section{Conclusion}

We have
developed a systematic analysis of a ground-state  atom undergoing a prescribed non-relativistic motion and coupled to the quantum 
electromagnetic field,  supposed to be initially in the vacuum state. 
We have assumed an harmonic motion of frequency $\omega_{\rm cm}.$ However, more general situations can be obtained from our formalism by 
Fourier decomposition. 

When $\omega_{\rm cm}$ is larger than the internal frequency $\omega_0,$
 the external motion drives a transition to an internal excited state, together with the emission of 
a single photon carrying the excess energy.  We have calculated the motion-induced excitation to first order in the perturbation provided by the dipolar Hamiltonian 
with a R\"ontgen correction. 
The  photons are emitted according to an angular distribution whose shape 
depends strongly on the ratio  $\omega_{\rm cm}/\omega_0.$
The total excitation rate is small since it scales as  $(v_m/c)^2.$ However, it increases with   $\omega_{\rm cm}/\omega_0,$
with  $\Gamma_{\rm{MIE}}\approx (2/3)(v_m/c)^2(\omega_{\rm cm}/\omega_0)^3 \Gamma_0$ for $\omega_{\rm cm}/\omega_0\gg 1.$ 

In the opposite case $\omega_{\rm cm}/\omega_0< 1,$ the leading effect is the parametric excitation of photon pairs to second order in the dipolar Hamiltonian. 
 Our approach provides a more fundamental
perspective into the DCE, which is usually  considered for macroscopic bodies with the help of constitutive equations and boundary conditions.
We have shown that several properties of the DCE can be explained at the atomic scale. For instance, the dependence of the total emission rate on the oscillation frequency for small compact objects is already obtained within our atomic model. 

Another important example is provided by an oscillating 
 plane mirror. We have modelled the material medium as a collection of ground-state atoms. We have assumed destructive interference 
 along the emission directions violating  the translational symmetry parallel to the mirror.
 In this way, we were able 
 to explain the main properties of the
emission angular distributions known for perfect metals,
although our description is clearly more appropriate for rarefied dielectric materials. This indicates that 
the DCE for different materials share common universal features which are already present at the  atomic level.

%
%\noindent
\begin{acknowledgments}
We thank   C. Farina, D. Dalvit, W. Wolff, R. Decca, R. L. Matos, H. Mirandola and G. Bi\'e for valuable discussions and
the Brazilian agencies National Council for Scientific and Technological Development (CNPq) and Rio de Janeiro Research Foundation (FAPERJ) for support. 
 P.A.M.N also acknowledges support from 
the Coordination for the Improvement of Higher Education Personnel (CAPES), the National Institute of Science and Technology Complex Fluids (INCT-FCx), and the S\~ao Paulo Research Foundation (FAPESP - 2014/50983-3).

\end{acknowledgments}

\appendix

\section{One-photon process\label{onephotonapp}}
%In this appendix we detail the matrix elements evaluation for one-photon process. 

In this Appendix, we detail the derivations of Sec.~\ref{mie} corresponding to the motion-induced excitation. We specify the main steps in order to recover Eq.~(\ref{dndomega}). 
For clarity, here we write the explicit time dependence of the operators in the interaction picture. 
We consider the electromagnetic field operators quantized in a 
 finite cubic box of side $L$, expressed in Gaussian units as 
\begin{eqnarray}
			\mathbf{\hat{E}}(r)\!\!&=&\!\!\sum_{\mathbf{k}\lambda} i \left(\frac{2\pi \hbar\omega}{L^3}\right)^{1/2}\!\!\!a_{\mathbf{k}\lambda}e^{ik \cdot r}\boldsymbol{\varepsilon}_{\mathbf{k}\lambda}  +\mbox{h.c.}  \label{e}  \\
			\mathbf{\hat{B}}(r)\!\!&=&\!\!\sum_{\mathbf{k}\lambda} i \left(\frac{2\pi \hbar \omega}{L^3}\right)^{1/2} a_{\mathbf{k}\lambda}e^{ik \cdot r} \mathbf{\hat{k}}\times\boldsymbol{\varepsilon}_{\mathbf{k}\lambda} + \mbox{h.c.} \, . \nonumber \\ \label{b}
					\end{eqnarray}
 h.c. denotes the hermitian conjugate of the series on the right-hand side, $\omega=|\mathbf{k}|c$, $k\cdot r=\mathbf{k}\cdot\mathbf{r}-\omega t$ and $\boldsymbol{\varepsilon}_{\mathbf{k}\lambda}$ are the polarization unit vectors. We first evaluate the matrix elements involving the Hamiltonian~(\ref{hint}) 
\begin{eqnarray}
&&\langle 1_{\mathbf{k}\lambda},e_s | H_{\rm int}(t) | g,0\rangle=i \left(\frac{2\pi \hbar\omega}{L^3}\right)^{1/2}\langle e_s |\hat{d}_j(t)|g \rangle  \cr\cr
&& \times	 e^{-ik\cdot r(t)}\left[(\boldsymbol{\varepsilon}_{\mathbf{k}\lambda})_j+\epsilon_{jmn}\frac{v_m(t)}{c}(\hat{\mathbf{k}}\times\boldsymbol{\varepsilon}_{\mathbf{k}\lambda} )_n \right]  \, .\label{eq:interactionmie}
\end{eqnarray}
We have used Einstein's convention for the summation over repeated indices and introduced the antisymmetric Levi-Civita tensor 
$\epsilon_{jmn}$ such that $\epsilon_{123}=1.$ We have
considered the general case in which there may be several excited states labeled by $s$.  The coupling~(\ref{eq:interactionmie}) contains simultaneously a static and a velocity-dependent contribution associated respectively to the electric and magnetic components of the Lorentz force. Substituting the above expression into Eq.~(\ref{psk}) we obtain for the probability of emission a sum of three terms related to contributions quadratic in the electric field, quadratic in the magnetic field, and bilinear in both fields. 

Note that the dipolar matrix elements can be written as $\langle e_s|\hat{d}_i(t)| g \rangle = \langle e_s|\hat{d}_i(0)| g \rangle e^{-i\omega_{s}t} $ in the interaction picture, where the frequency $\omega_s$ corresponds to the Bohr frequency between the ground state $g$ and the excited state $e_s$. Furthermore, when averaging over the possible atomic dipole configurations, by isotropy one obtains $  \overline{\langle e_s|\hat{d}_i(0)|g\rangle\langle g|\hat{d}_j(0)|e_s\rangle}=\delta_{ij}  |\langle e_s |    \mathbf{d}(0) | g \rangle|^2 /3.$ 

From now on, we detail specifically the contribution to the probability of emission $p_{s\mathbf{k}\lambda}^{(EE)}$ induced by terms quadratic in the electric field.
 We take the continuum limit $\sum_{\mathbf{k}} p_{s\mathbf{k}\lambda}\longrightarrow \frac{L^3}{8\pi^3}\int d^3\mathbf{k} p_{s\lambda}(\mathbf{k})$ and sum over the possible polarizations $\lambda$:
\begin{equation}
p_{s\mathbf{k}}^{(EE)}\!=\!\frac { |\langle e_s |    \mathbf{d}(0) | g \rangle|^2 \omega  }{6 \pi^2 \hbar}\!\int_{0}^{T}dtdt' e^{-i(\omega+\omega_{s})(t-t')}  e^{i \mathbf{k} \cdot( \mathbf{r}(t)-\mathbf{r}(t')))}
\end{equation}
As exposed in Sec.~\ref{mie}, in the non-relativistic regime one can treat perturbatively the external atomic motion described by Eq.~(\ref{cm}), i.e. one expands the complex exponential up to second order in the small parameter $ k a \ll 1 .$  It is then convenient to perform a variable change $(t,t') \rightarrow \left(\eta =\frac{1}{2}(t+t'),\tau=t-t' \right)$. Finally, one takes the long time limit, since one monitors the atomic emission over a time which is several orders of magnitude larger than the inverse of the atomic transition frequency. In this stationary limit, only resonant terms contribute to the emission process:
\begin{eqnarray}
		p_{s\mathbf{k}}^{(EE)} &=&\frac{  |\langle e_s |    \mathbf{d}(0) | g \rangle|^2 \omega T}{12\pi\hbar}(\mathbf{k}\cdot\boldsymbol{a})^2\delta(\omega+\omega_{s}-\omega_{\rm cm})
 \, ,\nonumber \\ \label{eefinal}
\end{eqnarray}
The contributions $p_{s\mathbf{k}}^{(EM)}$ and $p_{s\mathbf{k}}^{(MM)}$, respectively associated to terms bilinear in the electric and magnetic fields and quadratic in the magnetic field, may be obtained by following the same steps. Special care must be taken to work out consistently the perturbative expansion of the complex exponential as to obtain contributions on the order of $(v/c)^2.$ Finally, by integrating over the frequencies of emission 
 one obtains the angular distribution of emitted photons:
$d \Gamma_{\rm MIE} / d\Omega_k=(1/T) \sum\limits_{s}\int_{0}^{\infty}(p_{s\mathbf{k}}^{(EE)}+p_{s\mathbf{k}}^{(EM)}+p_{s\mathbf{k}}^{(MM)} ) k^2 d k.$ The resulting expression is  given by a sum of contributions of the form~(\ref{dndomega})  for   each excited state $e_s,$
with  $\omega_0$ replaced by $\omega_s$ and 
$\Gamma_0$ by the spontaneous emission rate $\Gamma_s$ between the excited state $e_s$ and the ground state $g.$

\section{Two-photon process\label{twophotons}}

In this section we present the derivation leading from  Eq.~(\ref{eq:coeffsck1k2})  to Eq.~(\ref{eq:angularspectrum}). First of all, from Eq.~(\ref{psit}) we see that the probability that two photons are created in the state $| 1_{\mathbf{k}_1  \lambda_1} 1_{\mathbf{k}_2 \lambda_2} \rangle$ is given in the stationary limit by
\begin{equation}
P_{\mathbf{k}_1\lambda_1,\mathbf{k}_2\lambda_2}=\lim_{t \rightarrow +\infty}|c_{\mathbf{k}_1 \lambda_1 \mathbf{k}_2  \lambda_2} (t)|^2 \, , \label{p}
\end{equation}
where  the limit physically means $t\gg 1/\Delta\omega.$
Now we use the identity (see for instance \cite{Landau})
\begin{equation}
\lim_{t \rightarrow +\infty} \frac{\sin^2(\Delta\omega\, t/2)}{\pi t\Delta\omega^2/2}=\delta(\Delta\omega) \, ,
\end{equation}
when substituting Eq.~(\ref{eq:coeffsck1k2}) into Eq.~(\ref{p}):
\begin{eqnarray}
&&\frac{P_{\mathbf{k}_1\lambda_1,\mathbf{k}_2\lambda_2}}{t}=\frac{2\pi^3\alpha_0^2\omega_1\omega_2v_m^2}{L^6c^2}\,\delta(\Delta\omega)\cr\cr &&\times\Bigg\{\mathbf{\hat{a}}\cdot\Bigg[\frac{c}{\omega_{\rm cm}}(\mathbf{k}_1+\mathbf{k}_2)(\boldsymbol{\epsilon}_{\mathbf{k}_1\lambda_1}\cdot\boldsymbol{\epsilon}_{\mathbf{k}_2\lambda_2})\cr\cr
&+&(\mathbf{\hat{k}}_1\times\boldsymbol{\epsilon}_{\mathbf{k}_1\lambda_1})
\times\boldsymbol{\epsilon}_{\mathbf{k}_2\lambda_2}+
(\mathbf{\hat{k}}_2\times\boldsymbol{\epsilon}_{\mathbf{k}_2\lambda_2})
\times\boldsymbol{\epsilon}_{\mathbf{k}_1\lambda_1}\Bigg]\Bigg\}^2 \, . 
\end{eqnarray}
The Dirac delta ensures the conservation of energy in the stationary regime. In the continuum limit we have
\begin{equation}
			\sum_{\mathbf{k}_1,\mathbf{k}_2}\longrightarrow \frac{L^6}{64\pi^6}\int d^3\mathbf{k}_1d^3\mathbf{k}_2 \, .
\end{equation}
and the probability becomes a density of probability given by
\begin{eqnarray}
&&\frac{P_{\lambda_1,\lambda_2}(\mathbf{k}_1,\mathbf{k}_2)}{t}=\frac{\alpha_0^2\omega_1\omega_2v_m^2}{32\pi^3c^2}\,\delta(\Delta\omega)\cr\cr &&\times\Bigg\{\mathbf{\hat{a}}\cdot\Bigg[\frac{c}{\omega_{\rm cm}}(\mathbf{k}_1+\mathbf{k}_2)(\boldsymbol{\epsilon}_{\mathbf{k}_1\lambda_1}\cdot\boldsymbol{\epsilon}_{\mathbf{k}_2\lambda_2})\cr\cr
&+&(\mathbf{\hat{k}}_1\times\boldsymbol{\epsilon}_{\mathbf{k}_1\lambda_1})
\times\boldsymbol{\epsilon}_{\mathbf{k}_2\lambda_2}+
(\mathbf{\hat{k}}_2\times\boldsymbol{\epsilon}_{\mathbf{k}_2\lambda_2})
\times\boldsymbol{\epsilon}_{\mathbf{k}_1\lambda_1}\Bigg]\Bigg\}^2 \, . \label{p12cont}
\end{eqnarray}
In order to obtain the photon production rate, we  integrate out one of the photons in the pair:
\begin{equation}
P_{\lambda}(\mathbf{k})=\sum_{\lambda_2}\int d^3\mathbf{k}_2P_{\lambda,\lambda_2}(\mathbf{k},\mathbf{k}_2) \, .
\end{equation}
Performing the Fourier space integration in spherical coordinates, we obtain from Eq.~(\ref{p12cont})
\begin{eqnarray}
&&\frac{P_{\lambda}(\mathbf{k})}{t}=\frac{\alpha_0^2\omega(\omega_{\rm cm}-\omega)^3v_m^2}{32\pi^3 c^5}\sum_{\lambda_2}\int d\Omega_{\mathbf{k}_2} \cr\cr
&&\Bigg\{\mathbf{\hat{a}}\cdot\Bigg[\frac{c}{\omega_{\rm cm}}\left[\mathbf{k}+\left(\frac{\omega_{\rm cm}}{c}-k\right)\hat{\mathbf{k}}_2\right](\boldsymbol{\epsilon}_{\mathbf{k}_1\lambda_1}\cdot\boldsymbol{\epsilon}_{\mathbf{k}_2\lambda_2})\cr\cr
&+&(\mathbf{\hat{k}}_1\times\boldsymbol{\epsilon}_{\mathbf{k}_1\lambda_1})
\times\boldsymbol{\epsilon}_{\mathbf{k}_2\lambda_2}+
(\mathbf{\hat{k}}_2\times\boldsymbol{\epsilon}_{\mathbf{k}_2\lambda_2})
\times\boldsymbol{\epsilon}_{\mathbf{k}_1\lambda_1}\Bigg]\Bigg\}^2 \, . \label{p1}
\end{eqnarray}

The angular integrals can be readily evaluated from symmetry considerations by relating them with averages over all spatial directions. 
Let us then analyze each type of  integral  required for the evaluation of Eq.~(\ref{p1}).
Firstly, 
\begin{equation}
\sum_{\lambda_2}\int d\Omega_{\mathbf{k}_2} (\boldsymbol{\epsilon}_{\mathbf{k}\lambda}\cdot\boldsymbol{\epsilon}_{\mathbf{k}_2\lambda_2})^2=4\pi\boldsymbol{\epsilon}_i^{\mathbf{k}\lambda}
\boldsymbol{\epsilon}_j^{\mathbf{k}\lambda}\sum_{\lambda_2}\overline{\boldsymbol{\epsilon}_i^{\mathbf{k}_2\lambda_2}\boldsymbol{\epsilon}_j^{\mathbf{k}_2\lambda_2}} \, , \label{p1a}
\end{equation}
where we employed Einstein summation convention. From symmetry, the tensor
obtained by averaging over all directions 
 in the r.-h.-s. of Eq.~(\ref{p1a}) 
 must be isotropic since we have summed over polarizations \cite{isotropic}:
\begin{equation}
\sum_{\lambda_2}\overline{\boldsymbol{\epsilon}_i^{\mathbf{k}_2\lambda_2}\boldsymbol{\epsilon}_j^{\mathbf{k}_2\lambda_2}}=C\delta_{ij}=\frac{2}{3}\delta_{ij} \, , \label{epsilonepsilon}
\end{equation}
where the constant $C$ were determined by contracting the indexes $i$ and $j$ on
both sides of the equation. 
Then, there is a term proportional to $\sum_{\lambda_2}\overline{\mathbf{\hat{k}}_{2i}\boldsymbol{\epsilon}_j^{\mathbf{k}_2\lambda_2}\boldsymbol{\epsilon}_m^{\mathbf{k}_2\lambda_2}}$ which must be proportional to 
the only isotropic  tensor of rank 3, 
$\epsilon_{ijm}.$
 However, the latter is
antisymmetric in the exchange $j\leftrightarrow m,$ while the former is symmetric, hence this term must vanish.
 There are also terms proportional to $\sum_{\lambda_2}\overline{k_{2i}\boldsymbol{\epsilon}_j^{\mathbf{k}_2\lambda_2}}$. These must be proportional to $\delta_{ij}$ and then vanish upon contraction of the indexes. 
 The most difficult integral we must deal with is proportional to
\begin{equation}
\sum_{\lambda_2}\overline{k_{2m}k_{2n}\boldsymbol{\epsilon}_i^{\mathbf{k}_2\lambda_2}\boldsymbol{\epsilon}_j^{\mathbf{k}_2\lambda_2}}=C_1\delta_{ij}\delta_{mn}+C_2(\delta_{im}\delta_{jn}+\delta_{in}\delta_{jm}) \label{kkepsilonepsilon} \, , 
\end{equation}
where we used the most general form of a 4-rank isotropic tensor 
symmetric upon the change $i\leftrightarrow j.$   Imposing that a contraction of $i$ with $m$ vanishes
while a contraction of $n$ with $m$ yields $2\delta_{ij}/3,$ we obtain the system of equations
\begin{eqnarray}
C_1+4C_2&=&0 \cr\cr
3C_1+2C_2 &=& \frac{2}{3} \, ,
\end{eqnarray}
which yields $C_1=4/15$ and $C_2=-1/15$. 

 We may evaluate  (\ref{p1}) with the results obtained in the previous paragraphs. 
 In order to obtain Eq.~(\ref{eq:angularspectrum}), we must 
only relate the probability of creating a photon with the number of photons created, which is given by the relation
\begin{equation}
d\mathcal{N}_{\lambda}(\mathbf{k})=P_{\lambda}(\mathbf{k})d^3\mathbf{k}=\frac{P_{\lambda}(\mathbf{k})\omega^2}{c^3}d\omega d\Omega_{\mathbf{\hat{k}}}
\end{equation}
Defining the rate of photon production by $\Gamma=\mathcal{N}/t,$ we write the spectral rate of photon creation by solid angle as
\begin{equation}
\frac{d\Gamma_{\lambda}}{d\omega d\Omega_{\mathbf{\hat{k}}}}(\omega,\hat{\mathbf{k}})=\frac{\omega^2P_{\lambda}(\omega,\hat{\mathbf{k}})}{tc^3}. \label{gammap}
\end{equation}


\begin{thebibliography}{99}
\bibitem{FootNoteSectionIIclassical} 
Such approach is legitimate if the shape of the atomic wave-function in momentum space is almost unaffected by the atomic recoil, and if the energy stored in the external motion is much larger than the energy quantum of the trap. These conditions can be simultaneously met by considering, for instance, 
the atomic center of mass to be
 in a coherent state $|\alpha\rangle$ with amplitude $|\alpha| \gg 1.$
\bibitem{FootnoteSecIIb} {
In order to compute the emission rate to order  $(v/c)^2,$ it is sufficient to expand the interaction Hamiltonian to order $v/c$ as in (\ref{hint}). 
}
\bibitem{secondorder} Up to order $(v/c)^2$ all processes involve the absorption of  a single  quantum
 of vibration $\hbar \omega){\rm cm}$ associated to the atomic center-of-mass. In second order of perturbation theory we have two-photon processes and by energy conservation we have in the stationary regime $\omega_{\rm cm}=\omega_{0s}+\omega_1+\omega_2$, where $\omega_{0s}$ is the atomic transition frequency to state $|s\rangle$ and $\omega_1$, $\omega_2$ are the frequencies of the produced photons. This can be immediately generalized for any order of perturbation theory and thus we conclude that up to order $(v/c)^2$ motion-induced excitation only happens if $\omega_{\rm cm}>\omega_{0s}$ for some sate $|s\rangle$ different from the ground state.
\bibitem{comment_comparison}  Note that the results for the sphere and the atom are not expected to coincide, since the magnetic dipole moment of a small metallic sphere at rest is comparable with its electric dipole moment.
\end{thebibliography}
\end{document}